\documentclass[journal]{IEEEtran}

\ifCLASSINFOpdf
\else
   \usepackage[dvips]{graphicx}
\fi
\usepackage{url}

\usepackage{algpseudocode}
\usepackage{algorithm}
\usepackage{graphicx}
\usepackage{ wasysym }
\usepackage{setspace}

\begin{document}

\title{Multi-Talker MVDR Beamforming Based on Extended Complex Gaussian Mixture Model}

\author{Hangting~Chen,
	Pengyuan~Zhang,~\IEEEmembership{Member,~IEEE,}
	and~Yonghong~Yan,~\IEEEmembership{Member,~IEEE,}
\thanks{Hangting Chen, Pengyuan Zhang and Yonghong Yan are with Key Laboratory of Speech Acoustics and Content Understanding, Institute of Acoustics, Chinese Academy of Sciences, Beijing 100190, China, email: chenhangting@hccl.ioa.ac.cn, zhangpengyuan@hccl.ioa.ac.cn, yanyonghong@hccl.ioa.ac.cn }
\thanks{Pengyuan Zhang is the corresponding author.}
}

\maketitle

\begin{abstract}
In this letter, we present a novel multi-talker minimum variance distortionless response (MVDR) beamforming as the front-end of an automatic speech recognition (ASR) system in a dinner party scenario. The CHiME-5 dataset is selected to evaluate our proposal for overlapping multi-talker scenario with severe noise. A detailed study on beamforming is conducted based on the proposed extended complex Gaussian mixture model (CGMM) integrated with various speech separation and speech enhancement masks. Three main changes are made to adopt the original CGMM-based MVDR for the multi-talker scenario. First, the number of Gaussian distributions is extended to $3$ with an additional inference speaker model. Second, the mixture coefficients are introduced as a supervisor to generate more elaborate masks and avoid the permutation problems. Moreover, we reorganize the MVDR and mask-based speech separation to achieve both noise reduction and target speaker extraction. With the official baseline ASR back-end, our front-end algorithm gained an absolute WER reduction of $13.87\%$ compared with the baseline front-end. 
\end{abstract}

\vspace{-0.1cm}

\IEEEpeerreviewmaketitle

\vspace{-0.3cm}
\section{Introduction}
\label{sec:intro}

After deep learning has been introduced in automatic speech recognition (ASR) \cite{dahl2012context-dependent}\cite{hinton2012deep}, modern systems have already achieved considerable recognition accuracies on the clean close-talk data \cite{xiong2016achieving}. However, speaker overlap, noise, and reverberation remains some of the biggest challenges. The CHiME challenges are targeting distant multiple microphone speech recognition in everyday listening environments \cite{barker2015the}. In the recent CHiME-5, the speech data is recorded in a party scenario, which presents extreme speech overlap and unrestrained speaking style \cite{barker2018the}. As a result, the word error rate (WER) of the official baseline system was $81.1\%$.

Beamforming strengthens a signal in a specific direction while attenuating it in other directions \cite{veen1988beamformings}, resulting in the reduction of noise and reverberation with controllable speech signal distortion \cite{spriet2004spatially}. The minimum variance distortionless response (MVDR) beamforming minimizes the power of the output signal while keep the original speech signal undistorted \cite{Cohen2010Speech}. However, the spatial correlation matrices of noise/speech signals are required for its implementation. A common method to obtain the correlation matrix is from time-frequency masks, which is averaged (or maximized) along all channels \cite{erdogan2016improved}. Recently, deep neural networks combined with ideal ratio masks (IRMs) have shown to generate considerably more elaborate separation masks \cite{erdogan2016improved}\cite{xiao2017on}\cite{heymann2016neural}. Several training losses, such as regression of IRM and indirect mapping (IM), were proposed and compared in \cite{Sun2017Multiple}. Additionally, network architectures like progressively learning training are more effective than fully-connected feed-forward networks \cite{gao2016snr-based}\cite{Gao2018Densely}. On the other hand, the complex Gaussian mixture model (CGMM) proposed in \cite{Higuchi2017Online} can generate masks as well as correlation matrices without any pairwise training samples. Some research attempts have been made to compare \cite{higuchi2018frame-by-frame} and integrate the neural network-based and CGMM-based methods \cite{nakatani2017integrating}.

In a cocktail party, however, beamforming cannot separate speaker sources owing to the close gathering of people in most scenarios. In this letter, a novel extended CGMM-based MVDR framework, which simultaneously removes inference speaker as well as noise, is designed to overcome the difficulties of speech overlap and noise in CHiME-5.

Our main contributions are two-fold. First, an extended CGMM algorithm is proposed by unifying speech separation (SS) and speech enhancement (SE) into one MVDR framework. Note that an extended CGMM algorithm was also proposed in the CHiME-5 submission \cite{CHiME_2018_paper_du}, but our algorithm does not require other front-end techniques or suffer from permutation problem. This simple, flexible algorithm integrated with single-channel masks is derived from the recent CGMM-based MVDR algorithm with $3$ main changes. An ablation study of proposed components is presented to make a detailed analysis of the algorithm. Second, different masks are compared to gain insight on the further improvement of this method. Our front-end, MVDR based on the extend CGMM integrated with single-channel SS and SE masks achieved a WER of $67.23\%$ under the baseline ASR, indicating the best front-end performance for Rank-A in CHiME-5.

\vspace{-0.25cm}
\section{CGMM-based MVDR Beamforming}
\label{sec:intro_cgmm_mvdr}

This section briefly formulates the problem of multiple sources and multichannel observation, and the corresponding CGMM-based approach to estimate speech masks, which follows the notation in \cite{Higuchi2017Online}.

\vspace{-0.25cm}
\subsection{MVDR Beamforming}
\label{subsec:mvdr}

Let $k \in {1,...,K}$ be a source index, $m \in {1,...,M}$ be a microphone index, $f \in {1,...,F}$ be a frequency index after the short-time Fourier transform (STFT). The signals received by the $m$th microphone in the time-frequency domain, $y_m$, can be represented as,
\vspace{-0.15cm}
\begin{equation}
\label{equ:mvdr1}
y_m(f,t)=\sum_{k} h_m^{(k)}(f)s^{(k)}(f,t)+n_m(f,t),
\vspace{-0.15cm}
\end{equation}
where $s^{(k)}$, $n_m$ and $h_m^{(k)}$ denote the $k$-th source signal, the noise received by $m$-th microphone, the impulse response between the $k$-th source and the $m$-th microphone, respectively. The Equation (\ref{equ:mvdr1}) can be rewritten in a vector notation as,
\vspace{-0.15cm}
\begin{equation}
\mathbf{y}(f,t)=\sum_{k} \mathbf{h}^{(k)}(f)s^{(k)}(f,t)+\mathbf{n}(f,t).
\vspace{-0.15cm}
\end{equation}

A beamforming based on the linear model applies a linear filter $\mathbf{w}^{(k)}_f$ on the microphone array signal $\mathbf{y}$,
\vspace{-0.15cm}
\begin{equation}
\label{equ:est_s}
\hat{s}^{(k)}(f,t)=\mathbf{w}^{(k)^{H}}_f\mathbf{y}(f,t).
\vspace{-0.15cm}
\end{equation}
The MVDR minimizes the variance of $\hat{s}^{(k)}$ subject to $\mathbf{w}^{(k)^{H}}\mathbf{h}^{(k)}(f)=1$. Thus the $\mathbf{w}^{(k)}_f$ can be written as,
\vspace{-0.15cm}
\begin{equation}
\label{equ:cal_w}
\mathbf{w}^{(k)}_f=\frac{R_f^{(y)^{(-1)}}\mathbf{h}_f^{(k)}}{\mathbf{h}_f^{(k)^H}R_f^{(y)^{-1}}\mathbf{h}_f^{(k)}},
\vspace{-0.15cm}
\end{equation}
where $R_f^{(y)}$ denotes the covariance matrix of noisy signal, calculated by
$R_f^{(y)}=\frac{1}{T}\sum_{t}\mathbf{y}_{f,t}\mathbf{y}_{f,t}^H$. The room transfer function $\mathbf{h}_f^{(k)}$ can be estimated as the eigenvector corresponding to the maximum eigenvalue of $R_f^{(k)}$, and it is determined by
\vspace{-0.15cm}
\begin{equation}
\label{equ:spatial_corr}
R_f^{(k)}=\frac{1}{\sum_{t}\lambda_{f,t}^{(k)}}\sum_{t}\lambda_{f,t}^{(k)}\mathbf{y}_{f,t}\mathbf{y}_{f,t}^H,
\vspace{-0.15cm}
\end{equation}
where $\lambda_{f,t}^{(k)}$ represents how the target source $k$ dominates the time-frequency bin. In practice, IRMs can be regarded equal to $\lambda_{f,t}^{(k)}$, which serves an external knowledge of target source to implement the MVDR. The algorithm of Equation (\ref{equ:est_s}), (\ref{equ:cal_w}) and (\ref{equ:spatial_corr}) is denoted as MVDR($X$,$masks$) with multichannel STFT spectrum and target source masks as input.

\vspace{-0.25cm}
\subsection{Complex Gaussian Mixture Model}
\label{subsec:cgmm}

The CGMM has been proposed to cluster time-frequency bins into acoustic sources as well as noise, then to estimate spatial correlation matrices. Based on the observation model, the multichannel signal from a source can be modeled in a complex Gaussian distribution,
\vspace{-0.15cm}
\begin{equation}
\label{equ:single_gua}
\mathbf{y}_{f,t}|k \sim \mathcal{N}_c(0,\phi_{f,t}^{(k)} \mathbf{R}_f^{(k)}),
\vspace{-0.15cm}
\end{equation}
where $k$ indicates the condition of the specific source, $\phi_{f,t}^{(k)}$ is the variance of signal in the time-frequency bin and $\mathbf{R}_f^{(k)}$ corresponds to $\mathbf{h}_f^{(k)}\mathbf{h}_f^{(k)^H}$. Introducing the mixture model for different sources, the multichannel observed signal can be modeled as,
\vspace{-0.15cm}
\begin{equation}
\mathbf{y}_{f,t} \sim \sum_{k}\mathcal{N}_c(0,\phi_{f,t}^{(k)} \mathbf{R}_f^{(k)}),
\vspace{-0.15cm}
\end{equation}
where the mixture coefficients are set to $1$. Thus, the log-likelihood function maximized by the Expectation-Maximization (EM) algorithm is defined as,
\vspace{-0.15cm}
\begin{equation}
\label{equ:loghood}
log\ p(y|\Theta)=\sum_{f,t}log\sum_{k}\mathcal{N}_c(\mathbf{y}_{f,t} | 0,\phi_{f,t}^{(k)} \mathbf{R}_f^{(k)}),
\vspace{-0.15cm}
\end{equation}
In the E step, the masks of each source can be estimated as,
\vspace{-0.15cm}
\begin{equation}
\label{equ:E1}
\lambda_{f,t}^{(k)} \leftarrow \frac{\mathcal{N}_c(\mathbf{y}_{f,t}|0,\phi_{f,t}^{(k)} \mathbf{R}_f^{(k)})}{\sum_{k}\mathcal{N}_c(\mathbf{y}_{f,t}|0,\phi_{f,t}^{(k)} \mathbf{R}_f^{(k)})}.
\vspace{-0.15cm}
\end{equation}
In the M step, the parameters are updated as follows,
\vspace{-0.15cm}
\begin{equation}
\label{equ:M1}
\phi_{f,t}^{(k)} \leftarrow \frac{1}{M}tr(\mathbf{y}_{f,t}\mathbf{y}_{f,t}^H \mathbf{R}_f^{(k)^{-1}}),
\end{equation}
\vspace{-0.45cm}
\begin{equation}
\label{equ:M2}
\mathbf{R}_f^{(k)} \leftarrow \frac{1}{\sum_{t}\lambda_{f,t}^{(k)}}\sum_{t}\frac{\lambda_{f,t}^{(k)}}{\phi_{f,t}^{(k)}}\mathbf{y}_{f,t}\mathbf{y}_{f,t}^H.
\end{equation}

\vspace{-0.25cm}
\section{Proposed Methods}
\label{sec:proposal}

\begin{algorithm}[t]
	\label{alg:alg1}
	\caption{The Beamforming Framework}
	\hspace*{0.02in} {$X$:}
	Multichannel STFT spectrum\\
	\hspace*{0.02in} {$Y$:}
	STFT spectrum after beamforming 
	\begin{algorithmic}[1]
		\State Split the whole spectrum into $blocks$, whose length is $T_{block}(ms)$
		\For{$block$}
		\If{Applying multichannel SS masks}
		\State \underline{Early-SS($X$,$masks$)}
		\EndIf
		\If{Applying multichannel SE masks}
		\State \underline{Early-SE($X$,$masks$)}
		\EndIf
		\If{Delay-and-sum}
		\State \underline{BeamformIT($X$)}
		\ElsIf{MVDR}
		\If{CGMM}
		\If{2CGMM}
		\State \underline{2CGMM-ini($X$,$masks$,$Prior$)}/
		\State	\underline{2CGMM-w/o-ini($X$)}
		\ElsIf{3CGMM}
		\State \underline{3CGMM($X$,$masks$,$Prior$)}
		\EndIf
		\EndIf
		\State \underline{MVDR($X$,$masks$)}
		\If{Applying single-channel SS masks}
		\State\underline{Later-SS($Y$,$masks$)}
		\EndIf
		\EndIf
		\EndFor
		\State Concatenate processed $blocks$ into the STFT spectrum
		\State \Return STFT spectrum
	\end{algorithmic}
\end{algorithm}

In this section, we mainly focus on a multi-talker CGMM-based MVDR algorithm. To implement this algorithm, a knowledge of masks of the target speaker and noise is required at first. The method of obtaining these masks is described in Section \ref{sec:exp}. 

\vspace{-0.25cm}
\subsection{Reorganize CGMM-based MVDR}
\label{subsec:3cgmm_alg}

The talking style of natural conversation causes extremely overlapped speech in CHiME-5, considered as one of the biggest challenges \cite{sun2019a}. Unfortunately, the speakers in CHiME-5 can not be modeled simply as separated signal sources. In most cases, a multichannel microphone array receives the unidirectional signal from the gathered speakers, resulting in the failure of blind speaker separation. In other words, utilizing the MVDR algorithm by considering the target and inference speakers as different sources can not solve the problem of speech overlap.

To separate different speakers, a speaker separation mask is first required, which can be generated by neural network-based methods like \cite{sun2019a}\cite{wu2019improved}. A straightforward approach, named Early-SS/Later-SS in the letter, is to reduce noise and reverberation with MVDR and apply the SS mask before/after the beamforming. 

Another concern is the iteration of the 3CGMM when using target, inference and noise masks simultaneously. In our practice, the masks can be iterated directly using EM algorithm in Equation (\ref{equ:E1}), (\ref{equ:M1}), (\ref{equ:M2}) with a number of $k$ equal to $3$.

\vspace{-0.25cm}
\subsection{Introduce Mixture Coefficients}
\label{subsec:mixture_coeffts}

\begin{figure}[t]
	\label{fig:masks}
	\centerline{\includegraphics[width=\columnwidth]{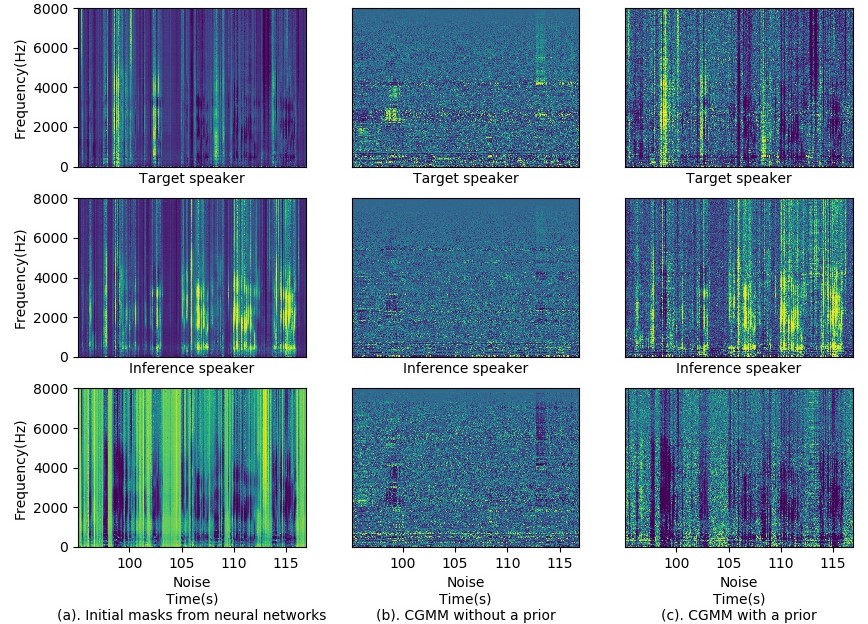}}
	\caption{The masks of the target speaker, inference speaker and noise for session S02 from $95s$ to $116s$, generated by (a) neural networks, (b) CGMM without a prior, (c) CGMM with a prior. The target speaker is P05 and the inference speakers are P06, P07, P08.}
	\vspace{-0.5cm}
\end{figure}

The EM algorithm for iterating the CGMM depends on the initialization of the $k$th source's mask $\lambda_k$. With a mask estimated by the neural networks, the CGMM's parameters can be initialized in a more accurate way. Nevertheless, in our CHiME-5 experiments, initialization (sophisticated neural network-based initialization or identity/random correlation matrices) leads to negligible divergence of final results. As shown in Figure 1(b), after sufficient iterations, the clustering pattern collapsed without contrast. The reason might be small divergence of the probability density computed by Gaussian models in Equation (\ref{equ:single_gua}). For audios with high signal-to-noise ratios (SNRs), the likelihood from different sources displays high contrast, leading to distinguishable masks. For low-SNR audios, the likelihood is low in all source CGMMs, resulting in indistinguishable estimated masks.

To solve the failure of clustering, a prior probability is introduced in Equation (\ref{equ:loghood}), 
\vspace{-0.15cm}
\begin{equation}
\label{equ:mixed_loghood}
log\ p(y|\Theta)=\sum_{f,t}log\sum_{k}\alpha_{f,t}^{(k)}\mathcal{N}_c(\mathbf{y}_{f,t} | 0,\phi_{f,t}^{(k)} \mathbf{R}_f^{(k)}),
\vspace{-0.15cm}
\end{equation}
which served as mixture coefficients in the CGMM. Different from \cite{Higuchi2017Online}, the mixture coefficients here are not scalar. It shares the same dimension with the STFT spectrum and is fixed among iterations. Only the update rules of Equation (\ref{equ:E1}) are rewritten as follows,
\vspace{-0.15cm}
\begin{equation}
\label{equ:E1new}
\lambda_{f,t}^{(k)} \leftarrow \frac{\alpha_{f,t}^{(k)}\mathcal{N}_c(\mathbf{y}_{f,t}|0,\phi_{f,t}^{(k)} \mathbf{R}_f^{(k)})}{\sum_{k}\alpha_{f,t}^{(k)}\mathcal{N}_c(\mathbf{y}_{f,t}|0,\phi_{f,t}^{(k)} \mathbf{R}_f^{(k)})}.
\vspace{-0.15cm}
\end{equation}
Note that using the mixture coefficients, the permutation problem are unobserved in most cases. The prior serves as a supervisor so that the only thing Gaussian models (Equation (\ref{equ:single_gua})) need to do is to add perturbation on it.  

\vspace{-0.25cm}
\subsection{The Complete Algorithm}
\label{subsec:whole_alg}

An extended 3CGMM-based MVDR is proposed here to alienate the difficulties of speaker overlap and noise corruption in CHiME-5. Here $3$ main changes are made to adapt the algorithm for the multi-talker condition,

\begin{enumerate}
\item \underline{Prior=True}. Introduce the mixture coefficient as a prior for CGMM, which equals to the input mask.
\item \underline{3CGMM}. The extended CGMM algorithm which iterates the target speaker, inference speaker and noise masks simultaneously.
\item \underline{Early-SS/Later-SS}. The target speaker mask is applied before/after beamforming to ensure that the processed audio contains only the target speaker. The masks can be averaged multichannel masks or the ones after 3CGMM iteration.
\end{enumerate}

Other components of experiments are also listed for comparison,
\begin{enumerate}
\item \underline{Early-SE}. Apply the denosing mask on the spectrum.
\item \underline{BeamformIT}. The delay-and-sum beamforming with BeamformIT \cite{anguera2006robust}\cite{anguera2007acoustic}.
\item \underline{MVDR}. The MVDR beamforming in Section \ref{subsec:mvdr}.
\item \underline{2CGMM-ini/2CGMM-w/o-ini}. The 2CGMM algorithm described in Section \ref{subsec:cgmm} with/without initialized correlation matrices and masks.
\end{enumerate}

The entire setup is given in Algorithm 1. We underline the main changes with code names, which will be carefully studied in Section \ref{subsec:ablation_algorithm}.

\vspace{-0.25cm}
\section{Experiments and Results}
\label{sec:exp}

\vspace{-0.10cm}
\subsection{Experiment Settings}
\label{subsec:exp_sets}

\begin{table*}[!htb]
	\caption{The ablation study of the beamforming framework in algorithm 1}
	\label{tab:ablation_study}
	\centering
	\small
	\setlength{\tabcolsep}{3pt}
	\begin{tabular}{|c|c|l|c|}
		\hline
		System ID & Beamforming & Components of Algorithm 1 & WER(\%) \\
		\hline\hline
		A0 & BeamformIT & BeamformIT & $80.59$ \\
		\hline
		A1 & BeamformIT & BeamformIT+Early-SE(masks=RT) & $80.77$ \\
		\hline
		A2 & BeamformIT & BeamformIT+Early-SS(mask=SS1) & $\mathbf{75.83}$ \\
		\hline
		A3 & BeamformIT & BeamformIT+Early-SS(mask=SS1)+Early-SE(masks=RT) & $76.59$ \\
		\hline\hline
		B1 & MVDR & MVDR(masks=RT) & $80.74$ \\
		\hline
		B2 & MVDR & MVDR(masks=SS1) & $85.09$ \\
		\hline
		B3 & MVDR & MVDR(masks=RT)+Early-SS(mask=SS1) & $74.38$ \\
		\hline
		B4 & MVDR & MVDR(masks=RT)+Later-SS(mask=SS1) & $\mathbf{73.68}$ \\
		\hline\hline
		C0 & 2CGMM-MVDR & MVDR(masks=2CGMM-w/o-ini)+2CGMM-w/o-ini & $81.35$ \\
		\hline
		C1 & 2CGMM-MVDR & MVDR(masks=2CGMM-ini)+2CGMM-ini(masks=RT,Prior=false) & $81.49$ \\
		\hline
		C2 & 2CGMM-MVDR & MVDR(masks=2CGMM-ini)+2CGMM-ini(masks=RT,Prior=false)+Early-SS(mask=SS1)  & $77.27$ \\
		\hline
		C3 & 2CGMM-MVDR & MVDR(masks=2CGMM-ini)+2CGMM-ini(masks=RT,Prior=false)+Later-SS(mask=SS1) & $74.64$ \\
		\hline
		C4 & 2CGMM-MVDR & MVDR(masks=2CGMM-ini)+2CGMM-ini(masks=RT,Prior=true)+Later-SS(mask=SS1) & $\mathbf{73.14}$ \\
		\hline\hline
		D1 & 3CGMM-MVDR & MVDR(masks=3CGMM)+3CGMM(masks=RT\&SS1,Prior=false)+Later-SS(masks=SS1) & $75.05$ \\
		\hline
		D2 & 3CGMM-MVDR & MVDR(masks=3CGMM)+3CGMM(masks=RT\&SS1,Prior=true)+Later-SS(masks=SS1) & $70.66$ \\
		\hline
		D3 & 3CGMM-MVDR & MVDR(masks=3CGMM)+3CGMM(masks=RT\&SS1,Prior=true)+Later-SS(mask=3CGMM) & $\mathbf{67.90}$ \\
		\hline
	\end{tabular}
\vspace{-0.3cm}
\end{table*}

Our experiments adopted the official baseline of CHiME-5 as the back-end ASR module \cite{CHiME5_baseline}. It was trained using worn data and $100k$ randomly selected far-field utterances under an architecture of time-delay neural networks in lattice-free MMI criterion. The performance test was conducted on the development set (Single Device Track, constrained LM, ranking A) \cite{barker2018the}. The baseline WER was $80.59\%$ under our machines with BeamformIT (A0 in Table \ref{tab:ablation_study}).

For all the front-end experiments described in this letter, the STFT spectrum was extracted every $16ms$ over $32ms$ hamming windows. The length of each block processed by the MVDR beamforming was set to $T_{blokcks}=8208ms$ equal to $512$ frames. The number of iterations for 2CGMM/3CGMM was $10$, which could achieve an average real-time factor of $0.51$/$0.81$ with a $2.20$GHz CPU.

\vspace{-0.25cm}
\subsection{Speech Separation and Enhancement Models}
\label{subsec:ss_model}

\begin{table}[!htb]
	\vspace{-0.3cm}
	\caption{The results of speaker separation models using BeamformIT}
	\label{tab:ss_models}
	\centering
	\small
	\setlength{\tabcolsep}{10pt}
	\begin{tabular}{|p{70pt}|p{75pt}|}
		\hline
		Approach & WER(\%) \\
		\hline
		Raw & $80.63$ \\
		\hline
		$1$-stage & $75.83$ \\
		\hline
		$2$-stage & $\mathbf{75.46}$ \\
		\hline
	\end{tabular}
\end{table}

Our speech separation model is speaker-dependent, i.e., the models were trained for each speaker in each session individually. The training data was simulated from non-overlapping parts of records. This feed-forward DNN had $3$ layers and $2048$ units in each. The inputs were extended to a context of $7$ frames. The same models were used for $2$-stage separation, but little improvement was observed in our experiments using BeamformIT and the baseline ASR back-end (Table \ref{tab:ss_models}). More details on the $2$-stage speech separation could be found in \cite{sun2019a}.

\begin{table}[!htb]
	\vspace{-0.3cm}
	\caption{The results of denoising approaches using BeamformIT}
	\label{tab:denoise_models}
	\centering
	\small
	\setlength{\tabcolsep}{10pt}
	\begin{tabular}{|p{75pt}|c|c|p{75pt}|}
		\hline
		Approach & PESQ & WER(\%) \\
		\hline
		Raw & $2.543$ & $\mathbf{80.63}$ \\
		\hline
		RT & $2.664$ & $80.77$ \\
		\hline
		PLT & $\mathbf{2.675}$ & $80.84$ \\
		\hline
	\end{tabular}
\vspace{-0.2cm}
\end{table}

Our speech enhancement model was trained using the original far-field records. Without additional front-end denoising modules, the training of the neural network was aimed at mapping the original far-field noisy speech mixed with the far-field noise to the original one. In practice, the SS model utilized regression training (RT) or progressively learning training (PLT) of IRM. The RT used the same architecture of the SS model, updated by the mean square loss of the IRM. The PLT architecture also adopted the fully-connected layers with an input extended to a context of $7$. It learned $5$ targets, and did not do post processing \cite{Gao2018Densely}. Without the presence of original clean data and due to the limitation of the single-channel denoising method, the network could not achieve any WER reduction but did promote the perceptual evaluation of speech quality (PESQ) on the simulated audios (Table \ref{tab:denoise_models}).
\vspace{-0.25cm}
\subsection{The Ablation Study of Algorithm 1}
\label{subsec:ablation_algorithm}

To analyze the beamforming framework in detail, we further conducted ablation studies of different collections of components in Algorithm 1 (Table \ref{tab:ablation_study}). The masks are marked as "SS1, RT" if they are from the 1-stage SS models and the RT models (Section \ref{subsec:ss_model}), and are marked as "2CGMM, 3CGMM" if they are equal to those in Equation (\ref{equ:E1})/(\ref{equ:E1new}) after iterations. 

From A0-A3 and B1-B2, the SS masks seemed to play a more positive role than the SE masks. The results of B2 indicated that the straight-forward using of SS masks in the MVDR algorithm could not remove inference speakers. However, an alternative way of applying the SS masks before/after MVDR beamforming dramatically improved the performance (B1-B4, C0-C3), and the Later-SS is more preferred (B3-B4, C2-C3). In contrast with MVDR (B4), the 2CGMM-MVDR (C3) as well as 3CGMM-MVDR (D1) did not exhibit superiority until a prior was introduced as mixture coefficients (C4, D2). The prior led to a more elaborate mask compared with the neural network-based mask and no pattern collapse occurred (Figure 1). At last, the initial SS masks were replaced by the ones after 3CGMM iterations (D3), leading to a further improvement of performance.

\subsection{Study of Different Masks}
\label{subsec:diff_masks}

\begin{table}[!htb]
	\caption{The results of the proposed 3CGMM-MVDR integrated with various masks}
	\label{tab:various_masks}
	\centering
	\small
	\setlength{\tabcolsep}{10pt}
	\begin{tabular}{|c|c|c|c|}
		\hline
		SS & SE & WER(\%) \\
		\hline
		$1$-stage & RT & $67.90$ \\
		\hline
		$1$-stage & PLT & $67.70$ \\
		\hline
		$2$-stage & RT & $67.45$ \\
		\hline
		$2$-stage & PLT & $\mathbf{67.23}$ \\
		\hline
	\end{tabular}
\end{table} 

Another study of the impact of different masks as the prior was conducted using various SS and SE masks on System D3 (Table \ref{tab:various_masks}). The experiments implied that a better prior mask led to a better performance, indicating that for further performance improvement, research should be conducted to obtain more accurate masks.


\vspace{-0.25cm}
\section{Conclusion}
\label{sec:conclusion}

In this letter, we propose a multi-talker MVDR beamforming based on an extended CGMM. A detailed study of the algorithm and the influence of different masks was conducted on the CHiME-5 dataset. Our best 3CGMM-MVDR system achieved a WER of $67.23\%$ without other front-end techniques compared with the $80.59\%$ WER of the official baseline system.

\bibliographystyle{IEEEtran}
\bibliography{refs}

\end{document}